# Revealing a mode interplay that controls second harmonic radiation in gold nanoantennas

*Jérémy Butet,[†,]\* Gabriel D. Bernasconi,[†] Marlène Petit,[§] Alexandre Bouhelier,[§] Chen Yan,[†]*

*Olivier J. F. Martin,[†] Benoît Cluzel[§], Olivier Demichel[§,]\**

[†] Nanophotonics and Metrology Laboratory (NAM),

Swiss Federal Institute of Technology Lausanne (EPFL),

1015, Lausanne, Switzerland.

[§] Laboratoire Interdisciplinaire Carnot de Bourgogne,

UMR 6303 CNRS-Université Bourgogne Franche-Comté,

21078 Dijon, France.

*Corresponding author E-mails: jeremy.butet@epfl.ch; olivier.demichel@u-bourgogne.fr

**ABSTRACT**: In this work, we investigate the generation of second harmonic light by gold nanorods and demonstrate that the collected nonlinear intensity depends upon a phase interplay between different modes available in the nanostructure. By recording the backward and forward emitted second harmonic signals from nanorods with various lengths, we find that the maximum nonlinear signal emitted in the forward and backward directions is not obtained for the same nanorod length. We confirm the experimental results with the help of full-wave computations done with a surface integral equation method. These observations are explained by the multipolar nature of the second harmonic emission, which emphasizes the role played by the relative phase between the second harmonic modes. Our findings are of a particular importance for the design of plasmonic nanostructures with controllable nonlinear emission and nonlinear plasmonic sensors as well as for the coherent control of harmonic generations in plasmonic nanostructures.

Motivated by different specific features, the study of nonlinear optical processes in plasmonic nanostructures has become a vivid field of research [1, 2]. First, the intrinsic nonlinear response of plasmonic materials enables the investigation of subtle nonlinear mechanisms associated with the surface [3, 4], the shape [2], the roughness [5, 6], and the symmetry [2] of plasmonic nanostructures. Second, local field enhancement associated with the plasmon resonances can boost nonlinear processes including second harmonic generation (SHG) [7-9], third harmonic generation [10-12], and nonlinear photoluminescence [13-16], such that these nonlinear signals provide indirect entry to the local field enhancement [17-19]. Third, the plasmonic modes associated with a nanostructure are the underlying framework upon which nonlinear processes can be built [20, 21]. It is only quite recently that the role played in nonlinear plasmonics by the interaction between the different modes available at the fundamental and harmonic frequencies has been recognized, leading to different multiresonant nanostructure designs that benefit from the interaction of several plasmonic modes at different frequencies [21-25]. This is especially the case for SHG where a dipolar excitation at the fundamental frequency produces a nonlinear signal which is essentially quadrupolar [26].

In this article, we shed new light on the interplay between the underlying modal structure supported by a plasmonic nanostructure and the corresponding SHG. Specifically, we show that the second harmonic emission from gold nanorods can be controlled using interferences between two modes excited at the second harmonic wavelength. By recording the backward and forward second harmonic signals emitted by plasmonic nanorods with different lengths at a fixed pump wavelength, we observed that the nonlinear signal emitted in opposite directions is maximum for different nanorod lengths, revealing surprising enhancement mechanisms that do not fit with the well-established relation between the nonlinear response and the plasmonic enhancement. We explain these experimental findings with the help of full-wave computations performed with a surface integral equation method (SIE); the numerical results emphasize the multipolar nature of the second harmonic emission and the associated interference effects [27]. These observations are interesting for the design of efficient second harmonic nanosources as well as for the development of nonlinear plasmonic sensing, which aims at probing small refractive index changes with the help of nonlinear plasmonic nanostructures.

The gold nanorods used in this work are fabricated on a glass coverslip with a standard electron beam lithography fabrication method [28]. The nanorods were designed with a width of 80 nm and a length varying from L = 90 nm to L = 160 nm with 10 nm steps. The SEM images reveal that the width is 75 ± 5nm and the nanorod lengths are between 15 nm and 30 nm shorter than the expected ones, see Fig. 1.The real nanorod lengths are indicated in the legend of Fig. 1(a). The thickness of the deposited gold layer is 50 nm. The linear optical response is obtained from dark-field spectroscopy. Figure 1(a) shows the evolution of the scattering spectrum for a series of gold nanorods whose lengths are reported close to the related spectrum. In order to perform the linear characterization in the same experimental conditions as those used for the nonlinear optical measurements, a drop of immersion oil is deposited on the top of the nanostructures. The nanorods are thus immersed in a homogenous medium with a refractive index of 1.5. The scattering spectra show that the localized surface plasmon resonance red-shifts as the nanorod length increases (from 650 nm for L = 75 nm to 890 nm for L = 138 nm), as reported in previous studies of the linear response of gold nanorods [29, 30]. One can note that resonances between the 110 and 120 nm long antennas seem to slightly blue-shift. However, this shift -also associated to a broadening of the resonances- appears to be not relevant because the shift is much smaller than the full width at half maximum of the resonance. To confirm numerically this behavior, we compute the scattering from nanorods with the similar dimensions with a SIE method [31]. The edges and corners of the rectangular nanorods are rounded with a radius of 5 nm to provide a more realistic model [32]. The nanorods are excited by a planewave polarized along the long axis and placed in a homogeneous medium with a refractive index of 1.5, similar to the refractive indices of both the glass substrate and immersion oil. The calculated localized surface plasmon resonance red-shifts from 710 nm for L = 90 nm to 990 nm for L = 160 nm. Calculated and measured spectra are in good agreement.

Having characterized the linear responses of the gold nanorods, we now turn our attention to their second harmonic responses. The second harmonic generation from colloidal gold nanorods has already been investigated with hyper-Rayleigh scattering using a collection at a right angle [33, 34] and for nanofabricated nanorods with a collection performed in the forward direction only [35, 36]. Note that the second harmonic intensity is also commonly collected in the backward direction using the same objective as that used for illumination [24, 37, 38]. All these works have emphasized the role played by localized surface plasmon resonances in the enhancement of the nonlinear emission. Inspired by a recent study of the second harmonic light emitted from AlGaAs dielectric nanoantennas [39], we choose in the present work to collect simultaneously the second harmonic intensity in both the forward and backward directions, as depicted in Fig. 2(a). A high numerical aperture oil-immersion objective (x60, NA 1.49) focuses on individual nanorods a 120 fs pulsed laser beam emitted at a wavelength of 820 nm. The pulse energy is kept constant to a value as low as 4.75 pJ, well below the damage threshold of the nanorods (the mean laser power is 380 µW). The intensity of the second harmonic emitted in the backward direction is collected by the same objective (episcopic objective), while the second harmonic intensity emitted in the forward direction is collected with a second oil-immersion objective (x100, NA 1.30 – diascopic objective). Two microscope objectives with distinct working distances are used for convenience of the alignment procedure. Figure 2(b) shows the second harmonic intensity collected in the forward and backward directions for nanorod lengths ranging from 75 nm to 138 nm (see the supplementary information for the experimental details – the corresponding measured spectra are shown in supporting information Fig. S1. Both SHG and nonlinear photoluminescence are observed). The error bars correspond to the sum of the noise level and the error in the estimation of the SHG signal (see supplementary information for more details). The intensities of the nonlinear signals in the forward and backward directions reach a maximum for two different nanorod lengths. According to the literature, this length-dependent signal is understood from the resonant excitation of a localized surface plasmon mode at the fundamental wavelength [1, 2]. This is confirmed by the experimental dark-field spectra, Fig. 1(a), indicating a resonant excitation of the longitudinal dipolar plasmon mode at the pump wavelength ($\lambda$ = 820 nm) for L = 110 nm and L = 120 nm. While the second harmonic intensity peaks at L = 110 nm in the forward direction, the maximum is reached for L = 120 nm in the backward direction. A similar behavior has been observed for a dozen of nanorod arrays, see the SHG from a second set of nanorods in Fig. 2(c), emphasizing the reproducibility of this observation. This observation cannot be inferred from the simple plasmonic enhancement of the nonlinear responses [36-38] and points towards a more complex mechanism.

To understand this behavior, we perform full-wave computations of the second harmonic response of the gold nanorod with a SIE method assuming a surface contribution to the nonlinearity [40] (see the supplementary information for the implementation details). The second harmonic spectra are evaluated for fundamental wavelengths spanning between 600 nm and 1100 nm and for nanorod lengths varying between 90 nm and 160 nm. The second harmonic intensity is integrated over a sphere in the far-field region (sphere radius 50 μm). As expected and illustrated in Fig. 3(a), the second harmonic emission is generally enhanced when the fundamental wavelength matches the dipolar longitudinal plasmon resonance. However, compared to the calculated scattering cross-sections displayed in Fig. 1(b) that exhibit a stronger linear response for larger antennas, the simulations in Fig. 3(a) indicate an enhanced SHG intensity for the small nanorods. Indeed, as the nanorod length decreases, the dipolar surface plasmon resonance blue shifts at the fundamental wavelength, augmenting the retardation effects and then the SHG [26]. For a fixed pump wavelength centered at 820 nm, the strongest enhancement of the nonlinear response occurs for 120 – 130 nm nanorod lengths.

We then calculate the emission pattern of the second harmonic wave for the different nanorod lengths considered in this work. Figure 3(b) shows the computed emission diagrams for four sizes (L = 100 nm, 110 nm, 120 nm, and 130 nm). In the supporting information, Fig. S3 gathers the computed patterns for all nanorod sizes. To compare these numerical results with experimental data, the computed second harmonic pattern is divided into two parts, corresponding to the second harmonic intensity integrated on the top hemisphere for the forward second harmonic intensity and that integrated on the bottom hemisphere for the backward second harmonic intensity. In the following, all the simulations are done with a fundamental wavelength of 820 nm. The results are shown in Fig. 3(c). We find that the length of the nanorods drastically influences the emission pattern. For L = 90 nm, the second harmonic is essentially forward emitted while for the longest nanorod, the diagram shows a stronger emission in the backward direction. We verified that this behavior is not modified by taking into account the different numerical apertures for the episcopic and diascopic objectives (see supporting information, Fig. S3). The evolution of the forward and backward second harmonic emissions in Fig. 3(c) is similar to the one observed in the experimental results shown in Fig. 2(b).

Before studying the underlying mechanisms of this effect, we recall that in the case of a normal planewave illumination of the nanorods, the SHG signal comes essentially from (i) a dipolar mode oriented along the illumination propagation direction and (ii) a quadrupolar mode oriented along the nanorods axis [2]. The dipolar second harmonic emission is thus due to the retardation effect induced by the fundamental wave: the phase variation of the fundamental wave across the rod thickness induces a second harmonic dipolar moment along the wavevector of the pump wave; see the left inset in Fig. 3(d). On the other hand, the quadrupolar second harmonic emission arises from the fundamental dipole excited along the nanorods axis; see the right inset in Fig. 3(d). In order to disentangle the contribution of these two modes, we perform a multipolar decomposition of the second harmonic emission, *i. e.* the second harmonic field is expressed using the vector spherical harmonics (VSHs) as [41]:

$$\mathbf{E}_{SHG} = \sum_{l=1}^{l_{max}} \sum_{m=-l}^{l} k_{SHG}^2 E_{l,m} \left[ a_{l,m} \mathbf{N}_{l,m} + b_{l,m} \mathbf{M}_{l,m} \right], \qquad (1)$$

where $a_{l,m}$ and $b_{l,m}$ are the complex expansion coefficients, $E_{l,m}$ is a normalization factor, $\mathbf{N}_{l,m}$ and $\mathbf{M}_{l,m}$ are the VSHs and $l_{max}$ is set to 8. The definition used for the VSHs and the normalization factor can be found in Ref. 41. The expansion coefficients are found by projecting the computed electromagnetic fields onto the VSHs at a distance of 10 μm from the nanorods. The amplitude of the three coefficients for the dipolar ($a_{1,m}$) or the five coefficients for the quadrupolar ($a_{2,m}$) second harmonic emissions are then summed up to determine the relative weight of the dipolar or quadrupolar emissions; orders higher than the quadrupole as well as the magnetic modes, related to the $b_{l,m}$ coefficients in Eq. (1), are found to be negligible. In Fig. 3(d), we observe that the second harmonic dipolar mode is the strongest contribution to the second harmonic emission, except for nanorod lengths between 110 nm and 130 nm, for which the longitudinal dipolar mode is resonantly excited at the pump wavelength (820 nm); see Fig. 1(b).

The evolution of the amplitudes of the dipolar and quadrupolar second harmonic emissions cannot explain the flip of the nonlinear responses discussed previously since these two contributions are symmetric with respect to the forward and backward directions as pictured by the insets in Fig. 3(d). The literature suggests that the interference between different second harmonic modes, and thus their relative phase, plays an important role in the observed emission patterns [42, 43]. Especially, it was shown that the phase induced by the fundamental dipolar mode allows controlling the forward and backward second harmonic emission [27, 42, 44]. For this reason, the multipolar analysis is further refined in Fig. 4. The projection of the second harmonic field on the VSHs reveals that the dipole contribution comes from the two expansion coefficients $a_{1,-1}$ and $a_{1,1}$ and that the quadrupole contribution comes from the three coefficients $a_{2,-2}$, $a_{2,2}$ and $a_{2,0}$, the remaining coefficients $a_{1,0}$, $a_{2,-1}$ and $a_{2,1}$ being negligible. Note that the coefficient values are dependent on the axis orientation, see Fig. 3(b). A careful analysis of the dipole vector and the quadrupolar matrix (see the supporting information), combined with symmetry considerations, provide the relation between the coefficients $a_{1,m}$ and $a_{2,m}$. For the dipole, the relations $a_{1,1}= a_{1,-1}$ and $a_{1,0}=0$ are obtained and, for the quadrupole, $|a_{2,-2}| = |a_{2,2}|$ and $\arg(a_{2,0}) = \arg(a_{2,-2})+\pi = \arg(a_{2,2})+\pi$, in agreement with the results shown in Fig. 4(a-b). This observation confirms that second harmonic emission corresponds to that of a transverse dipole and a longitudinal quadrupole. Since the phase between the quadrupoles coefficients is roughly constant (close to $\pi$) over the whole nanorod size range, see the green curves in Fig. 4(b), and the two dipole coefficients are in phase, only the relative phase between the dipole coefficients and the quadrupole coefficient $a_{2,0}$ is considered in the following. In Fig. 4(b), we observe that the phases of the dipolar and quadrupolar components do not evolve at the same rate (compare the blue and green curves), hinting that the flip in the second harmonic emission occurs when the phase difference between the two components reaches a specific value. Figure 4(c) shows the evolution of the phase difference between the dipole and quadrupole coefficients, revealing that the flip indeed occurs for a specific phase difference $\Delta\phi$ between 0.19 $\pi$ rad and 0.28 $\pi$ rad, corresponding to nanorod lengths between 115 nm and 120 nm. To confirm the role of this phase difference in the flip of the second harmonic emission, the VSHs associated to the dipolar mode have been added to those of the quadrupole mode with expansion coefficients of constant amplitudes but different relative phase $\Delta\phi$ as:

$$\mathbf{E}_{SHG} = \left[a_{1,-1}\mathbf{N}_{1,-1} + a_{1,1}\mathbf{N}_{1,1}\right] + e^{i\Delta\phi}\left[a_{2,-2}\mathbf{N}_{2,-2} + a_{2,0}\mathbf{N}_{2,0} + a_{2,2}\mathbf{N}_{2,2}\right], \qquad (2)$$

with $a_{1,-1} = a_{1,1} = 1$, $a_{2,-2} = a_{2,2} = -1$, and $a_{2,0} = 0.82$ to reproduce the dipolar and quadrupolar emissions. The result is shown in Fig. 4(d), where we observe that the direction flip occurs for a phase difference $\Delta\phi=0.15$ $\pi$ rad, a value very close to the one extracted from the scattering spectra in Fig. 4(c). Concerning the relation between $|a_{2,-2}|$ and $|a_{2,0}|$, we observe that the flip between backward and forward emission does not depend upon the VSH $N_{2,0}$ and thus of the amplitude of the $a_{2,0}$ coefficient (data not shown). This observation suggests that the flip in the nonlinear emission is due to fundamental symmetry relation between the VSHs. Although not required to obtain the flip in the nonlinear emission, the specific relation $|a_{2,0}| = 0.82|a_{2,-2}| = 0.82|a_{2,2}|$ reproduces a radiation pattern with the expected cylindrical symmetry around the $x$ axis (see the supporting information). The electric field associated with a VSH do not have a constant phase over the sphere (i.e. is a function of both $\theta$ and $\varphi$), leading to a complex interference process.

Beyond their importance for the fundamental understanding of the mechanisms that lead to second harmonic emission in plasmonic nanostructures, these observations are also very important for the development of nonlinear plasmonic sensing – the nonlinear analog of plasmonic sensing aiming at the detection of small refractive index changes using the nonlinear properties of plasmonic nanostructures [45-47]. Indeed, nonlinear signal collected in such an application, *i. e.* the second harmonic intensity, is not necessary maximized when the fundamental wavelength matches the scattering maximum. Furthermore, the modal interplay occurring between the dipolar and quadrupolar second harmonic emissions could be directly used for sensing. Finally, when designing a practical device for nonlinear sensing, it is crucial to collect the signal where it is strongest. Figure 5(a) shows an example of such an application where the forward and backward second harmonic intensities are plotted as a function of the refractive index of the surrounding medium. A nanorod length of 100 nm has been chosen as an example and the fundamental wavelength is 820 nm. The dependence upon the refractive index is identical to the one discussed previously considering the nanorod length influence, *i. e.* a flip of the nonlinear pattern is observed in Fig. 5(a). To quantify this flip, the ratio between the forward and backward second harmonic intensities is shown in Fig. 5(b). This ratio evolves between 0.8 and 1.9 for a refractive index of the surrounding medium between 1.5 and 2. In the region of the largest slope, the estimated sensitivity is 4.7 $RIU^{-1}$ (Refractive Index Unit). Although the comparison with other methods, especially the ones developed in the linear regime, is not straightforward [48, 49], the approach proposed here is a credible alternative to standard methods for in situ characterization during nonlinear and ultrafast optical measurements.

In conclusion, we have shown that the emission pattern of the SHG emitted from gold nanorods strongly depends on the length of the nanorods. We have experimentally demonstrated a different evolution of the second harmonic intensities recorded in the forward and backward directions using an original experimental configuration. Specifically, we have observed that the nonlinear emission in these two directions is not maximal for the same nanorod length, a behavior departing from the well-known plasmonic enhancement of the nonlinear response. Using a full-wave numerical method, we have clearly identified the underlying mechanisms that lead to the experimental observation and revealed that it stems from the evolution of the phase between the different multipoles involved in the second harmonic emission. This result indicates that, whilst the modal structure associated with a plasmonic nanostructure is key to understanding its spectral response, the dephasing between the different modes can govern fundamental properties, such as the emission direction and detected intensities. Since the recorded signal depends upon the experimental configuration of the detection, a special care is required to determine the nonlinear conversion rate and to design efficient nonlinear plasmonic nanostructures. Furthermore, we showed that the modal interplay occurring in the SHG from plasmonic nanostructures may be used to measure small refractive index changes [45-47]. These results can also open interesting directions for the dynamic control of harmonic generations in plasmonic nanostructures based on engineering the relative phase between the different modes [50, 51].

ACKNOWLEDGMENTS

This work has been performed in cooperation with the Labex ACTION Program (Contract No. ANR-11-LABX-0001-01). This research was performed while JB was at the Laboratoire Interdisciplinaire Carnot de Bourgogne as an invited scientist. JB, GDB, CY, and OJFM acknowledge funding from the Swiss National Science Foundation (SNSF, Projects 200020_153662 and 200021_162453). OJFM and AB acknowledge funding from the European Research Council (ERC-2015-AdG-695206 Nanofactory) and FP7/ 2007-2013 Grant Agreement No. 30677, respectively.

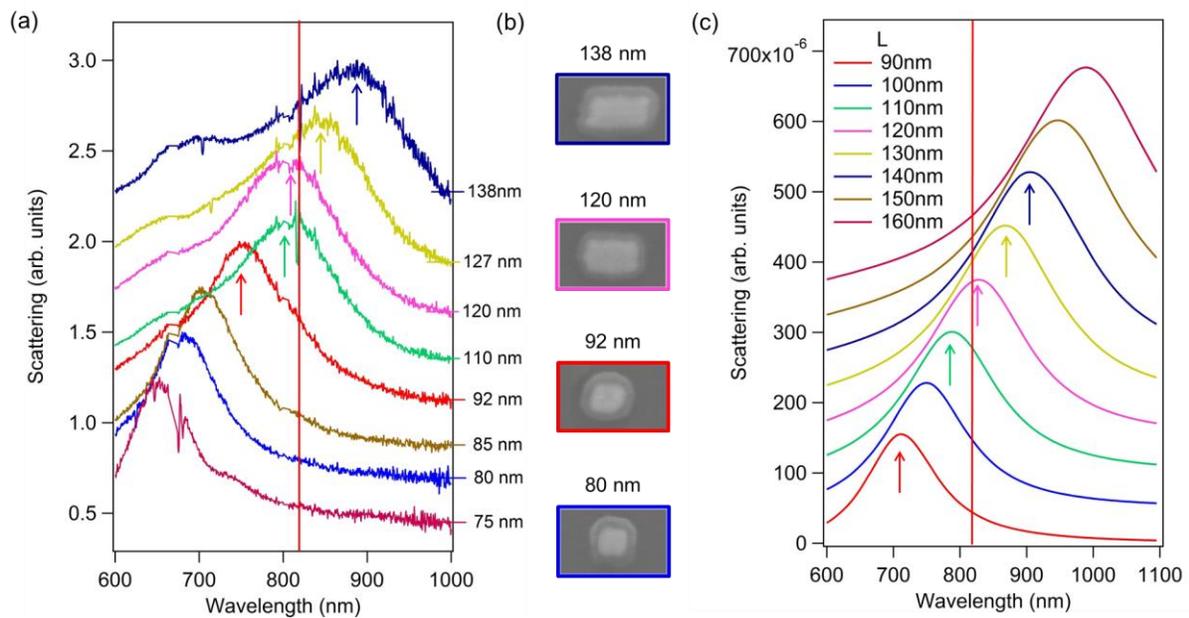

**Figure 1:** (a) Dark-field spectra of the gold nanorods with length ranging from 75 to 138 nm. (b) SEM images of four nanorods. (c) The scattering spectra of the nanorods, whose length vary from 90 to 160 nm, evaluated with a surface integral equation method. The nanorod widths and thicknesses are 75 nm and 50 nm, respectively. The spectra are shifted vertically for clarity.

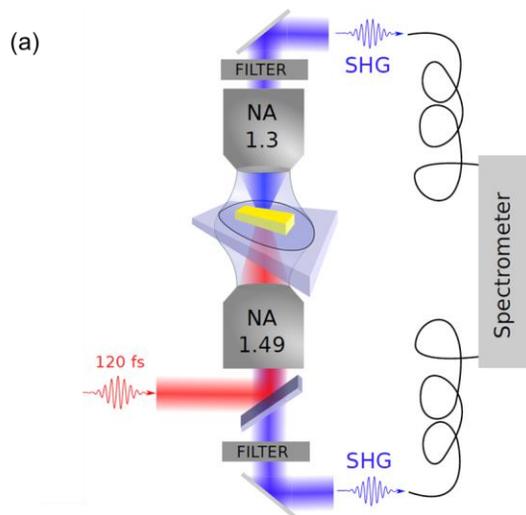

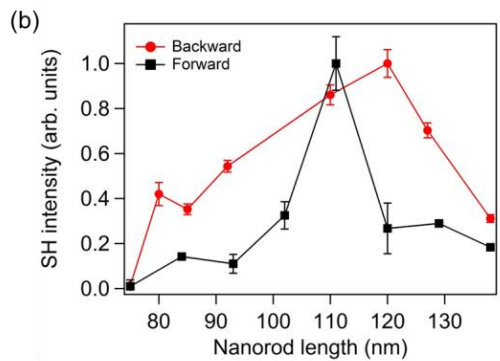

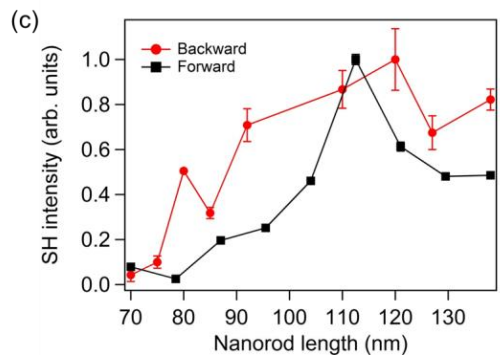

**Figure 2:** (a) Sketch of the experimental set-up used for the simultaneous detection of the forward and backward second harmonic light. (b) Forward (shown in black) and backward (shown in red) second harmonic intensities as functions of the nanorod length. The corresponding dark-field spectra are shown in Fig. 1(a). (c) Forward (shown in black) and backward (shown in red) second harmonic intensities for a second set of nanorods, showing the reproducibility of the experimental observations. The errorbars represent the error in estimating the SHG intensity.

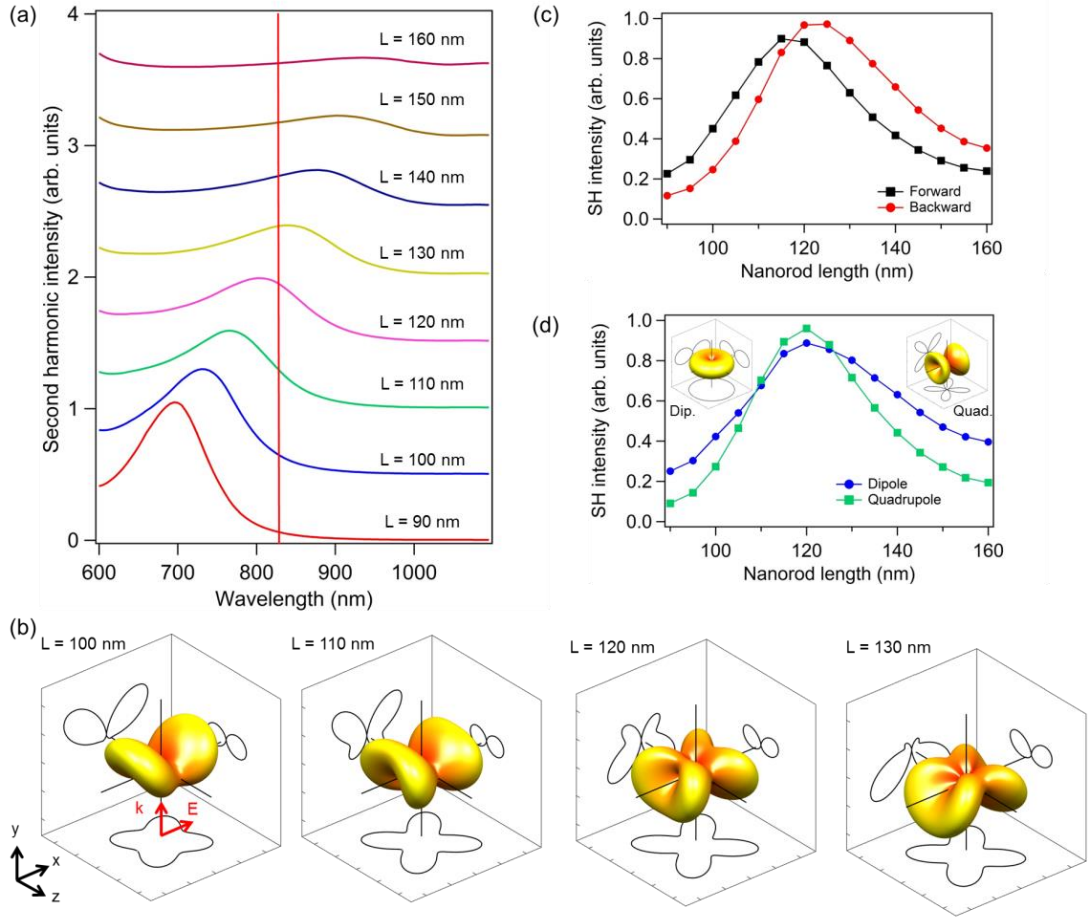

**Figure 3:** (a) Second harmonic intensity as a function of the illumination wavelength for nanorod lengths ranging from 90 nm to 160 nm. The spectra are shifted vertically for clarity. (b) Second harmonic emission patterns for nanorod lengths of 100 nm to 130 nm. The black solid lines correspond respectively to the pattern profiles in the planes x=0, y=0 and z=0. (c) Computed forward (shown in black) and backward (shown in red) second harmonic intensities as functions of the nanorod length. (d) Decomposition of the second harmonic intensity in dipolar (shown in blue) and quadrupolar (shown in green) emissions. The insets show the emission patterns of the dipolar and quadrupolar second harmonic modes. The fundamental wavelength is 820 nm, excepted for panel (a).

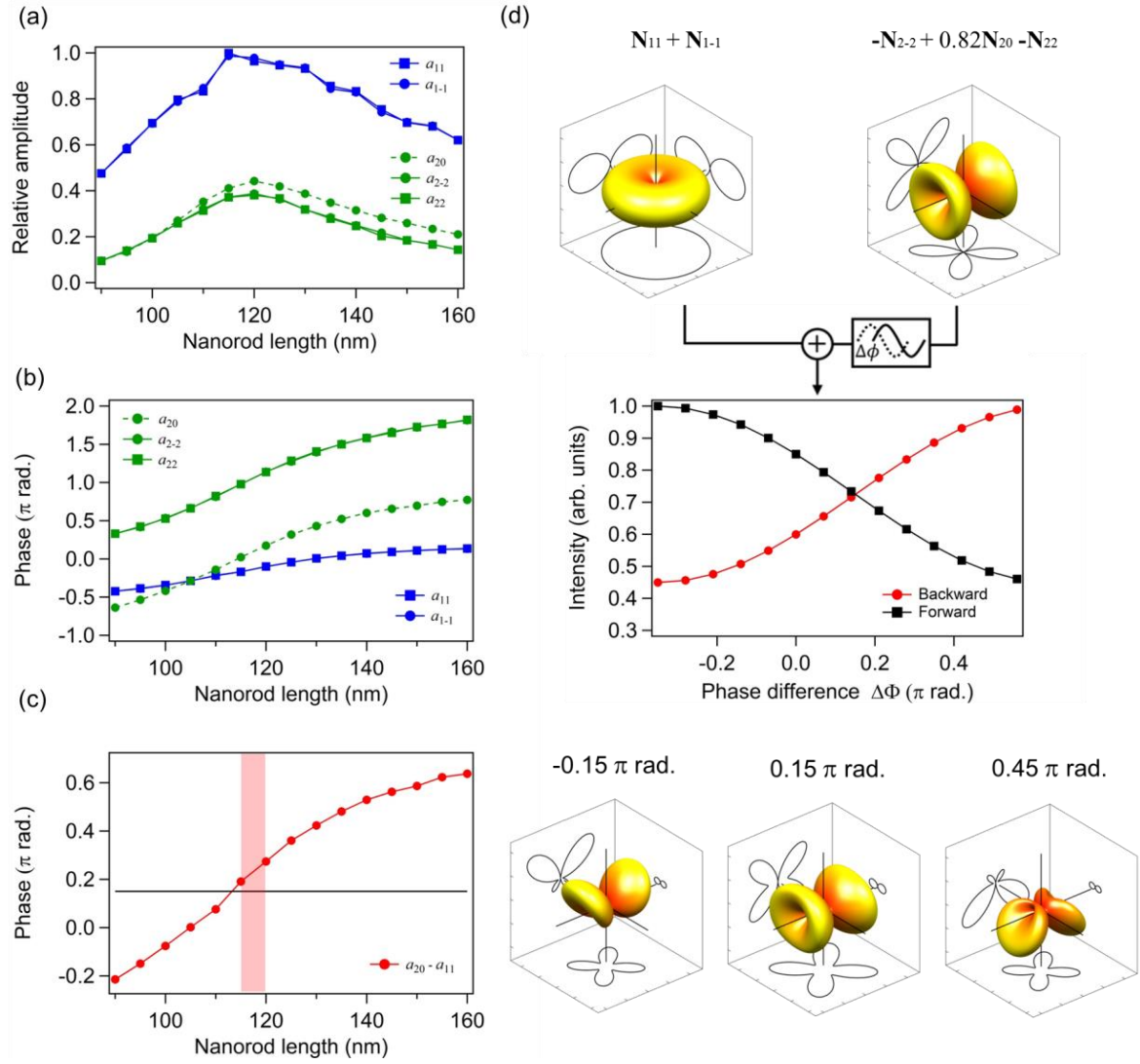

**Figure 4:** (a) Normalized amplitude and (b) phase of the expansion coefficients $a_{lm}$ for nanorod length ranging from 90 to 160 nm. (c) Phase difference between the dipolar coefficient $a_{11}$ and the quadrupolar coefficient $a_{20}$. (d) Evolution of the emission direction with the phase difference between $a_{11}$ and $a_{20}$. The second harmonic emission is computed as $\mathbf{E}_{SHG} = [\mathbf{N}_{1-1} + \mathbf{N}_{11}] + e^{i\Delta\phi}[-\mathbf{N}_{2-2} + 0.82\,\mathbf{N}_{20} - \mathbf{N}_{22}]$.

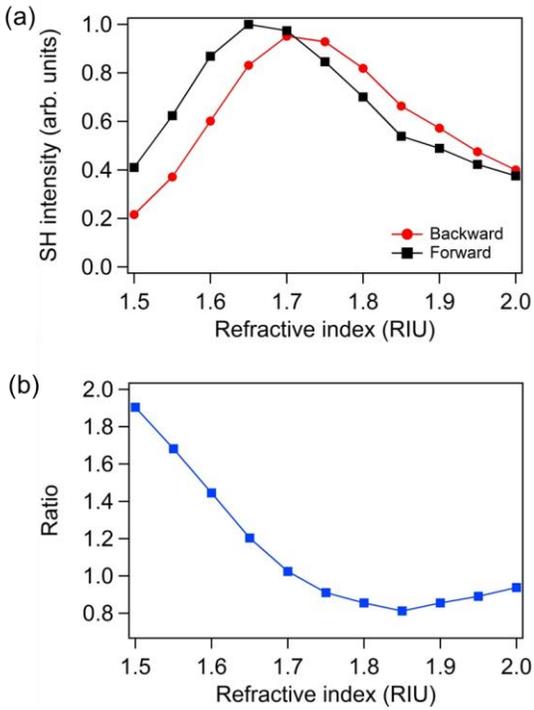

**Figure 5:** (a) Computed forward (shown in black) and backward (shown in red) second harmonic intensities as functions of the refractive index of the surrounding medium. (b) Ratio between the forward and backward second harmonic intensities as a function of the refractive index of the surrounding medium. The nanorod length is 100 nm and the fundamental wavelength is 820 nm.